\documentclass[11pt,a4paper]{article}
\usepackage{graphicx}
\begin{document}
\title{Study of the mixed Ising spins {\small $(\frac{1}{2},\frac{3}{2})$} in a random crystal field}

\author{L. Bahmad , A. Benyoussef \thanks{benyous@fsr.ac.ma}, and A. El Kenz \thanks{e-mail: elkenz@fsr.ac.ma} \\
\\
Facult\'e des Sciences, D\'epartement de Physique,  \\
Laboratoire de Magn\'etisme et Physique des Hautes Energies. \\
{\small  B.P. 1014, Rabat, Morocco}}

\date{}
\maketitle
\begin{abstract}
We study the magnetic properties of a mixed Ising ferrimagnetic system, in which the two interacting sublattices have spins $\sigma$, $(\pm 1/2)$ and spins $S$, $(\pm 3/2,\pm 1/2)$ in the presence of a random crystal field, with the mean field approach. The obtained results show the existence of some interesting phenomena, such as  the appearance of a new ferrimagnetic phase namely the partly ferrimagnetic phase $(m_{\sigma}=\frac{-1}{2},m_S=+1)$
and consequently the existence of three topologically different types of phase diagrams. The effect of increasing the exchange interaction parameter $J$, at very low temperature is investigated. The transitions shown in these phase diagrams are in good agreement with 
those obtained in the ground state case. \\
\end{abstract}



\section{Introduction}
The magnetic properties of two sublattices mixed spins $-1/2$ and spin $S > 1/2$ Ising system, with a crystal field interaction, have been recently studied both experimentally and theoretically. In fact, a crystal-field interaction effects on the transition temperature are investigated by several methods such as effective field theory \cite{KAN88,KAN90,1BENKK94,2BENKK94}, finite cluster approximation \cite{BENYA89}, mean field theory \cite{ABU01}, Migdal-Kadanof renormalisation group method \cite{BENYA90A} and cluster variational method \cite{TUC01}. However, there are some disagreements among those theoretical studies such as the existence of tricritical points and other features. Experimentally, the $MnNi(EDTA)-6H_2O$ complex has been shown to be an example of a mixed spin system \cite{DRI83}. Another interest towards the mixed spin Ising models can be related to the modelling of magnetic structures suitable for describing a ferrimagnetism of certain class of insulating materials. Indeed, these systems provide simple but interesting models to study molecular magnetic materials that are considered to be possibly useful materials for magneto-optical recordings. Furthermore, important advances have been made in the synthesis of two and three dimensional ferrimagnets, such as $2d$ organo-metallic ferrimagnets \cite{TAM93,OKA93}, $2d$ networks of the mixed-metal materials $[P(Phenyl)_4][MnCr(oxalate)_3]_n$ \cite{MAT96}. However, the possibility of many compensation points in a variety of ferrimagnetic systems has been clarified theoretically \cite{KAN95,JASC98}. \\
Since the ferrimagnetic order plays an important role in these materials, the investigation of ferrimagnetism in mixed spin systems has rapidly become a very rich field of research. As it is well known, these systems have less translational symmetry than their single-spin counterparts since they consist of two interpenetrating and non equivalent sublattices. From a theoretical point of view, many different methods have been employed in these studies. In particular, the mixed spin$-1/2$ and spin$-1$ Ising model has been solved exactly in special cases \cite{DAK98,OCT03}. On the other hand, the use of approximate methods such as mean field theory, free-fermion approximation, effective field theory, high-temperature series expansions, renormalisation group and Monte Carlo simulation, have revealed interesting results. \\
More recently, new magnetic properties and compensation behaviour are found for mixed spins in the presence of a crystal-field \cite{YAN07}. Such magnetic compensation behaviours were predicted by Nèel theory of ferrimagnetism \cite{BOBA95}, but not revealed in previous works. \\
On the other hand, the effective-field theory study, and multicritical behaviour, of the mixed spin Ising model with different anisotropies have revealed contradictions with earlier works obtained within the same theoretical framework. But the mean-field theory study based on the Bogoliubov inequality for the Gibbs free energy \cite{BOBA03}, it has been shown that in the presence of a single-ion anisotropy, the phase diagrams obtained exhibit a variety of multicritical points such as tricritical points and isolated critical points. \\
On the other hand, some recent experimental studies performed on various single-crystal samples $RVO_3$ ($R=La, Nd, Sm, Gd, Er,$ and $Y$) \cite{TUN07}, showed a temperature-induced and magnetization reversal. \\
The purpose of this work is to study, via mean field approximation (MFA), the influence of crystal-field disorder on the phase diagrams and magnetizations of a mixed-spin ferrimagnetic Ising system, in which the two interpenetrating sublattices have spins $\sigma=\pm 1/2$ and $S=\pm 3/2, \pm 1/2$. The most interesting result emerging from this study is the appearance of new types of phase diagrams, in the particular case of two-valued distribution of crystal-field. Consequently, three topologically different types of phase diagrams occur. 
This paper is organized as follows: in the section $2$, we introduce the model and give the details of the MFA. The ground-state phase diagram is discussed in section $3$. In section $4$ we present and discuss our results. Finally, section $5$ is devoted to summarizes and conclusions.

\section{Model and method.}
Since the MFA method neglects correlations between different spins, it is interesting to study the behaviour of complex spin systems, such as the ferrimagnetic mixed Ising models. The model we are studying consists of two interpenetrating sublattices. One sublattice has spins $\sigma$ assumed to take the values $\pm 1/2$, the other sublattice has spins $S$ that can take four values: $\pm 3/2$,  and $\pm 1/2$. The spins $S$ have only the spins $\sigma$ as nearest neighbours and vice versa.
The interaction between the spins $\sigma$ and $S$ is assumed to be an antiferromagnetic exchange. The Hamiltonian of this model is written as:
\begin{equation}
{\cal H}=J\sum_{<ij>}\sigma_i S_j+\sum_{i=1}^{N/2}\Delta_iS_i^2
 \label{hamilton}
\end{equation}
where $N$ is the total number of lattice sites. The exchange interaction parameter $J$ is assumed to be positive. The first summation is carried out only over nearest pairs of spins and $\Delta_i$ is a quenched random crystal field distributed according to the probability distribution \cite{BOCKS89,BAH07,BAH08,VIER01}:
\begin{equation}
{\cal P}(\Delta_i)=\frac{1}{2}[\delta(\Delta_i-\Delta(1+\alpha)) +\delta(\Delta_i-\Delta(1-\alpha))]
 \label{random}
\end{equation}
where $\alpha$ is a positive constant. \\
An analogous probability distribution has been used to investigate the critical behaviour of $^3He-^4He$ mixtures in random media (aerogel) modelled by the  spin$-1$ Blume-Capel model. In this model, the negative crystal-field value corresponds to the field at the pore-grain interface and the positive one is a bulk field that controls the concentration of  $^3He$ atoms \cite{MARI92,BUZ94,BRAN97}. \\

The variational principle based on the Gibbs-Bogoliubov inequality for the free energy per site is described by \cite{BOG47,FEY55,BIN92}:

\begin{equation}
{\cal F} \le \Phi = -T ln (Z_{0})+<{\cal H}-{\cal H}_0>_0
 \label{freenergy}
\end{equation}
Let us denote by $h_{\sigma}$ and $h_S$ the molecular fields associated with the order parameters $m_{\sigma}=<\sigma>_0$ and $m_S=<S>_0$, respectively, expressed as:
\begin{equation}
h_{\sigma}=J\sum_{j=1}^z<S_j>_0=zJm_S
 \label{hsigma}
\end{equation}
\begin{equation}
h_{S}=J\sum_{j=1}^z<\sigma>_0=zJm_{\sigma}
 \label{hs}
\end{equation}
where $z$ is the number of nearest neighbours and $<...>_0= \frac{Tr...exp(-\beta {\cal H}_0)}{exp(-\beta {\cal H}_0)}$ denotes the average value performed over the Hamiltonian ${\cal H}_0$.

The effective Hamiltonian of the system is given by:
\begin{equation}
{\cal H}_0=h_{\sigma}\sum_{i=1}^{N/2}\sigma_i +h_{S}\sum_{i=1}^{N/2}S_i+\sum_{i=1}^{N/2}\Delta_iS_i^2.
 \label{effectif}
\end{equation}
The partition function generated by the above Hamiltonian is :
\begin{eqnarray}
Z_{0}&=&Tr(exp(-\frac{{\cal H}_0}{T})) \nonumber \\
&=&(2cosh(\frac{\beta h_{\sigma}}{2}))^{N/2}(2 exp(\frac{-9\beta \Delta_i}{4}) cosh(\frac{3\beta h_{S}}{2})+2 exp(\frac{-\beta \Delta_i}{4}) cosh(\frac{\beta h_{S}}{2}))^{N/2}
 \label{partfunction}
\end{eqnarray}
where $T$ is absolute temperature and the Boltzmann's constant has been set to unity. 
The total free energy is given by:
\begin{eqnarray}
\Phi &= &\frac{NJz m_{\sigma}m_S}{2}-\frac{N h_{\sigma}m_{\sigma}}{2}-\frac{N h_S m_S}{2} \nonumber  \\
& & -NT \int Log(Z_{0}) {\cal P}( \Delta_i)d \Delta_i.
\label{ffreenergy}
\end{eqnarray}
After the integration over the probability distribution, the free energy per spin is given by:
\begin{eqnarray}
\frac{\Phi}{N}&=& \frac{NJz m_{\sigma}m_S}{2}-\frac{N h_{\sigma}m_{\sigma}}{2}-\frac{N h_S m_S}{2}-\frac{t}{2}[Log(2 cosh \frac{h_{\sigma}}{2t}) \nonumber  \\
& & +\frac{1}{2}(Log(2 exp(-\frac{9d(1+\alpha)}{4t})cosh(\frac{3h_S}{2t})+2exp(-\frac{d(1+\alpha)}{4t})cosh(\frac{h_S}{2t})) \nonumber   \\
& & Log(2 exp(-\frac{9d(1-\alpha)}{4t})cosh(\frac{3h_S}{2t})+2exp(-\frac{d(1-\alpha)}{4t})cosh(\frac{h_S}{2t}))
)] 
\label{reducedff}
\end{eqnarray}

In order to investigate the magnetizations of the system, the order parameters $m_{\sigma}$ and $m_S$
 are defined by minimizing the free energy.
Then, the mean-field equations of state are expressed as follows:
\begin{equation}
m_{\sigma}=-\frac{1}{2}tanh(\frac{z}{2t}Jm_S)
 \label{msigma}
\end{equation}
\begin{equation}
m_{S}=-\frac{1}{2}[\frac{A}{B} + \frac{C}{D}]
\label{ms}
\end{equation}
where,
\begin{eqnarray*}
A&=&3exp(-\frac{9d}{4t}(1+\alpha ) ) sinh(\frac{3 z m_{\sigma}}{2t})+exp(-\frac{d}{4t}(1+\alpha) ))sinh(\frac{z m_{\sigma}}{2t}) \\
B&=&2exp(-\frac{9d}{4t}(1+\alpha ) ) cosh(\frac{3 z m_{\sigma}}{2t})+2exp(-\frac{d}{4t}(1+\alpha) ))cosh(\frac{z m_{\sigma}}{2t}) \\
C&=&3exp(-\frac{9d}{4t}(1-\alpha ) ) sinh(\frac{3 z m_{\sigma}}{2t})+exp(-\frac{d}{4t}(1-\alpha) ))sinh(\frac{z m_{\sigma}}{2t}) \\
D&=&2exp(-\frac{9d}{4t}(1-\alpha ) ) cosh(\frac{3 z m_{\sigma}}{2t})+2exp(-\frac{d}{4t}(1-\alpha) ))cosh(\frac{z m_{\sigma}}{2t})
\end{eqnarray*}
In the above equations, and in all the following, $t$ and $d$ denote the reduced temperature $T/J$ and the reduced crystal field $\Delta/J$, respectively. The coordination number $z$ is set to be 4 (square lattice). \\
The solutions of the Eqs.~\ref{msigma}--\ref{ms} are not unique, the stable ones are those minimizing the free energy Eq.~\ref{ffreenergy}, while the others are the unstable ones. If the order parameters are continuous (discontinuous), the transitions are of second  (first) order.

\section{Ground state}
The ground state phase diagram of the system under investigation, according to the distribution of the crystal field law (Eq.~\ref{random}), is illustrated in Fig. 1. Indeed, for very low temperatures and depending on the values of $\alpha \ge 0$ and the reduced crystal field $d=\Delta /J$.
Eqs.~\ref{msigma}--\ref{ms} lead to three solutions: $(m_{\sigma}=-1/2,m_S=3/2)$, $(m_{\sigma}=\frac{-1}{2},m_S=\frac{1}{2})$, and $(m_{\sigma}=\frac{-1}{2},m_S=1)$. By comparing the energies for all possible configurations, we established the ground state phase diagram. This phase diagram is drawn in the reduced $(d,\alpha)$ plane. One can distinguish four cases: \\
1- For $\alpha=0$, a first order transition between the phase $(m_{\sigma}=-1/2,m_S=3/2)$ and the phase $(m_{\sigma}=\frac{-1}{2},m_S=\frac{1}{2})$ occurs at $d=+1$. \\
2- For $0 < \alpha < 1$, two first-order transition lines occur: in one hand between the phase $(m_{\sigma}=-1/2,m_S=3/2)$ and the phase $(m_{\sigma}=\frac{-1}{2},m_S=1)$ according to the equation line $d=z/(4(1+\alpha))$ and in the other hand between the phase $(m_{\sigma}=\frac{-1}{2},m_S=1)$and the phase $(m_{\sigma}=\frac{-1}{2},m_S=\frac{1}{2})$  according to the equation line $d=z/(4(1-\alpha))$. \\
\begin{figure}[!ht]
\begin{center}
\includegraphics[angle=-90,width=0.75\textwidth]{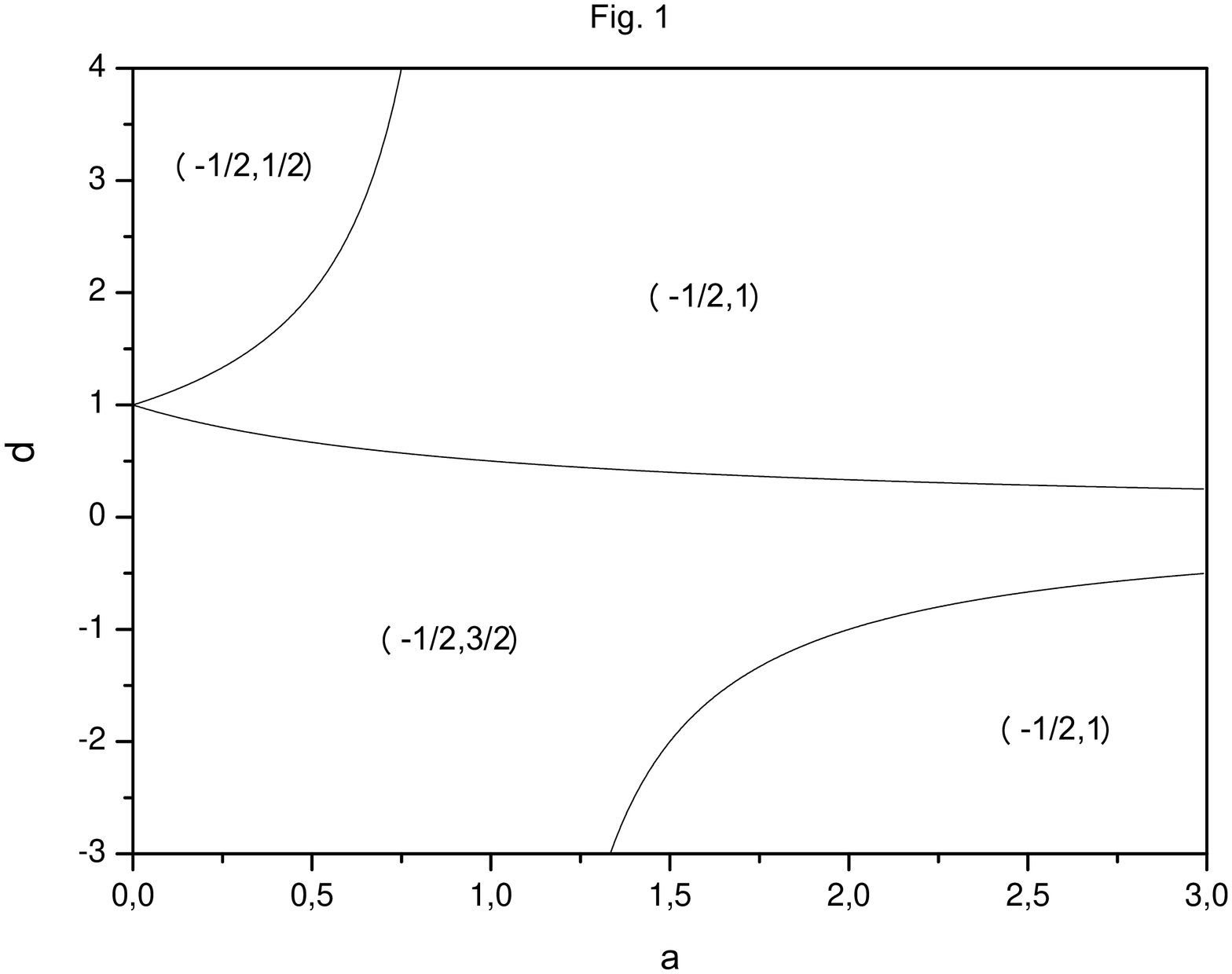}
\caption{The ground state phase diagram established in the $(d=\Delta /J,\alpha)$ plane. $(-1/2,1/2)$, $(-1/2,3/2)$ and $(-1/2,1)$ are the only stable phases for very low temperatures.}
\end{center}
\end{figure}
3- For $\alpha=1$, a first order transition between the phase $(m_{\sigma}=\frac{-1}{2},m_S=\frac{3}{2})$ and the ferrimagnetic phase $(m_{\sigma}=\frac{-1}{2},m_S=1)$ appears at $d=\frac{1}{2}$. \\
4- For $\alpha > 1$, two first-order transition lines between the $(m_{\sigma}=\frac{-1}{2},m_S=1)$ phase and the  $(m_{\sigma}=\frac{-1}{2},m_S=\frac{3}{2})$ phase and the $(m_{\sigma}=\frac{-1}{2},m_S=\frac{3}{2})$ phase and the $(m_{\sigma}=\frac{-1}{2},m_S=1)$ phase occur at $d=z/(4(1-\alpha))$ and $d=z/(4(1+\alpha))$, respectively. \\
It is worth to note that for $d=0$, only the phase $(m_{\sigma}=\frac{-1}{2},m_S=\frac{3}{2})$ is stable at $T=0 K$. For a higher temperature, the phase diagrams for different values of $\alpha$ and $d$ will be discussed in all the following.

\section{Phase diagrams and discussions}
\subsection{Phase diagrams}
A detailed discussion dealing with finite temperature phase diagrams is discussed in this section. For this purpose, we solve numerically the Eqs.~\ref{msigma},~\ref{ms} and ~\ref{ffreenergy}. A rich variety of phase transitions is observed both when varying $\alpha$ in the $(t_c=T_c/J,d=\Delta /J)$ plane, and $d$ in the $(t_c,\alpha)$ plane. Indeed, the critical temperature is plotted as a function of $d$ for $\alpha =0$ (Fig. 2a), $\alpha =0.5$ (Fig. 2b), $\alpha =1$ ( Fig. 2c) and $\alpha =2$ (Fig. 2d). 
\begin{figure}[!ht]
\begin{center}
\includegraphics[angle=-90,width=0.55\textwidth]{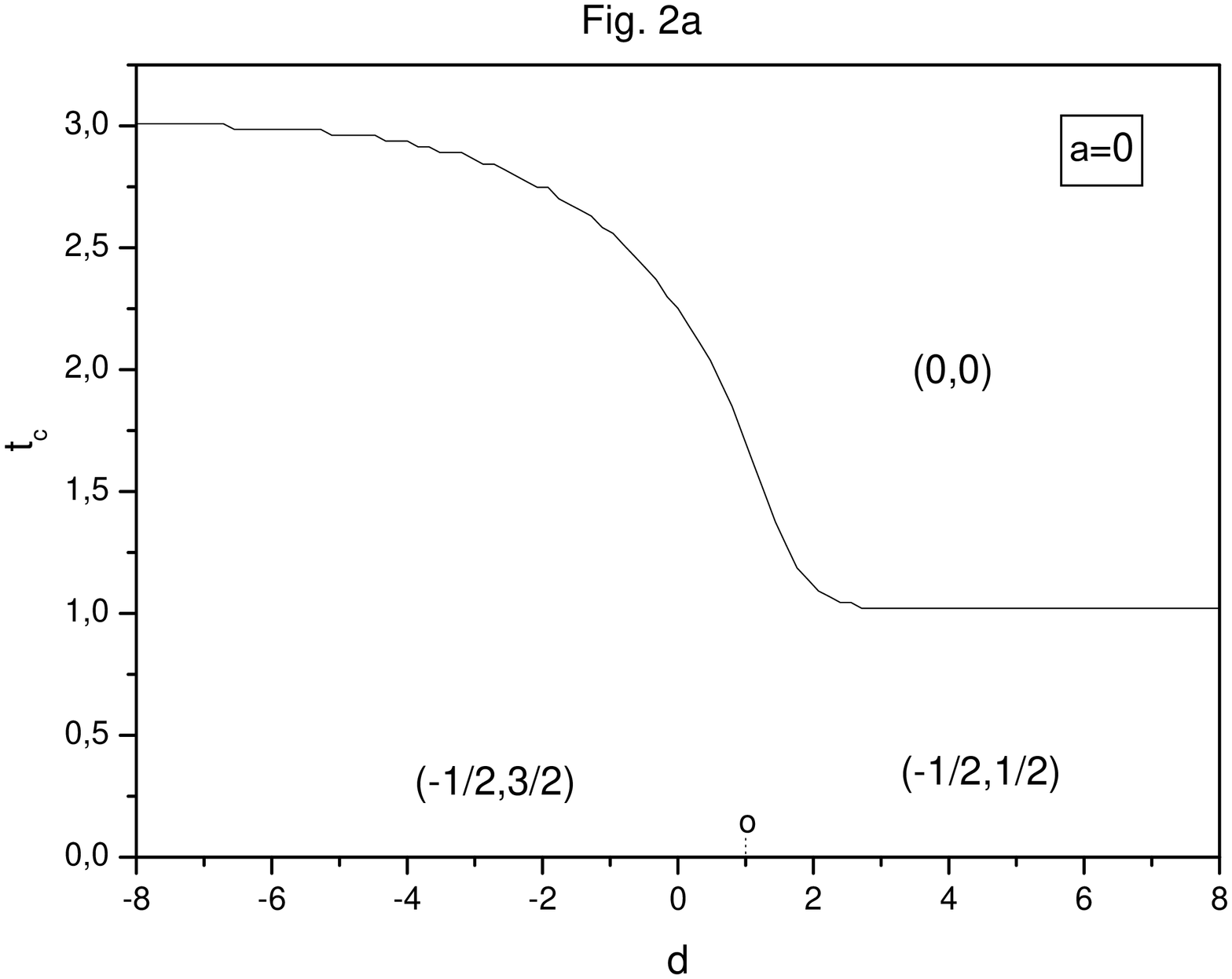}\qquad
\includegraphics[angle=-90,width=0.55\textwidth]{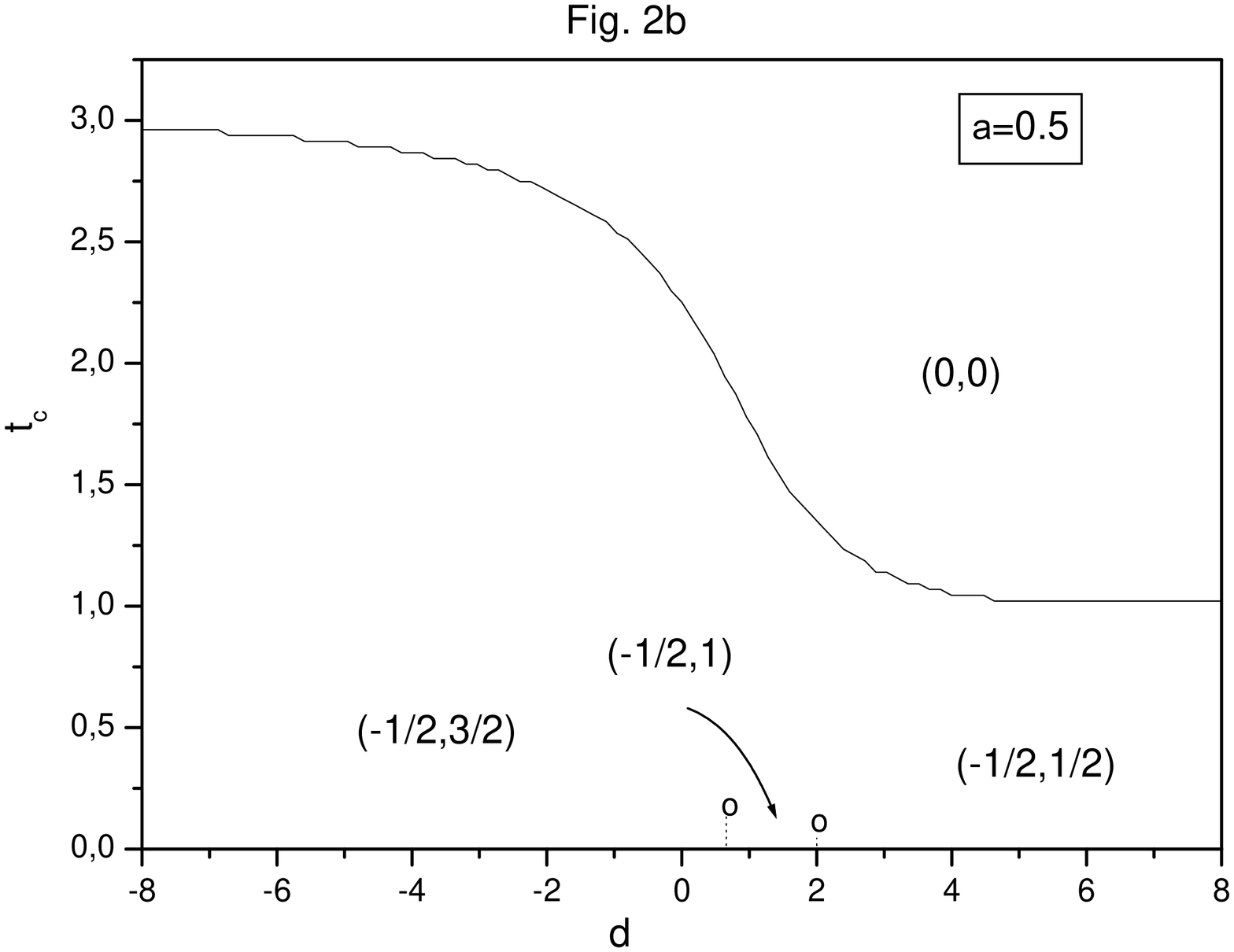}
\includegraphics[angle=-90,width=0.55\textwidth]{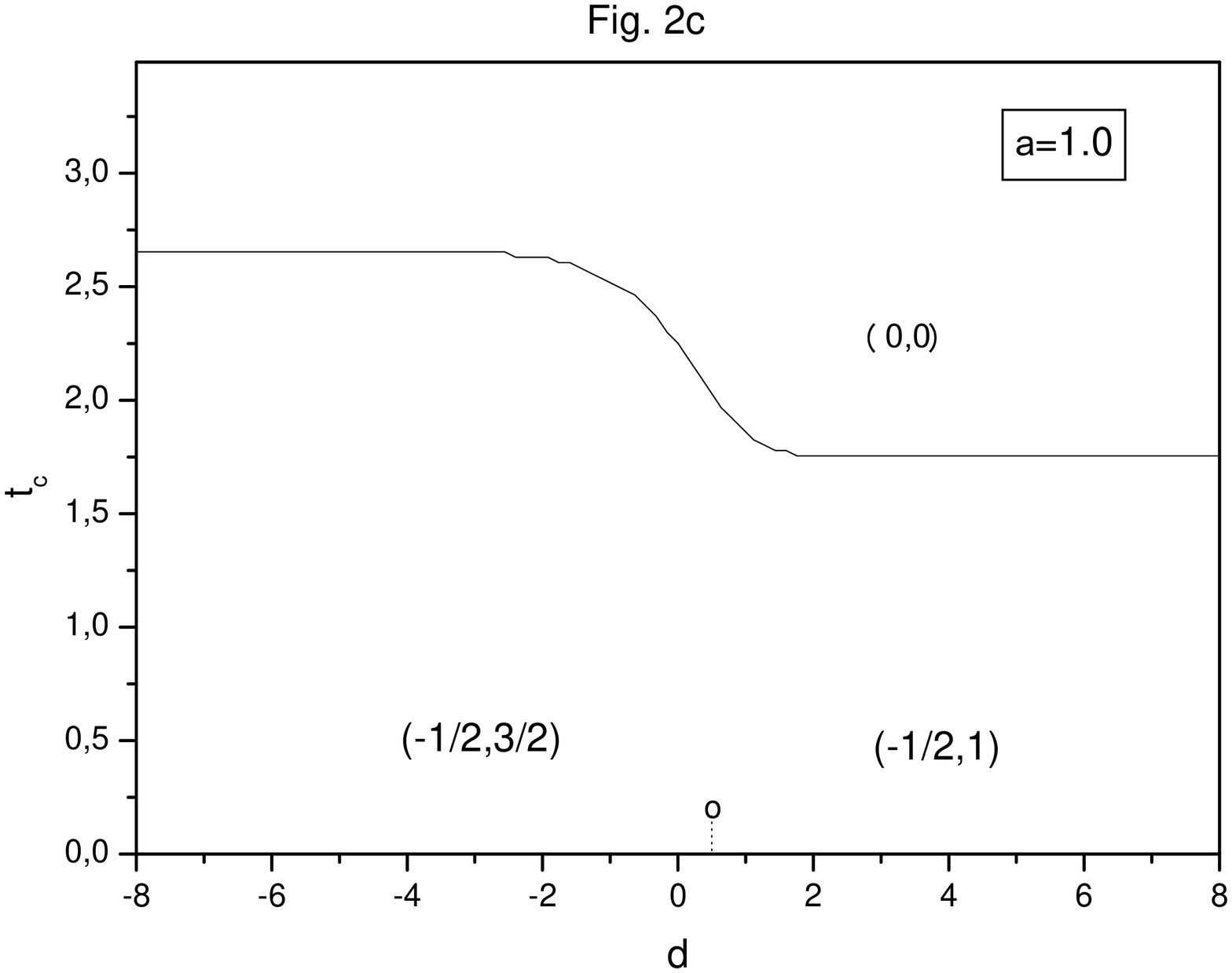}\qquad
\includegraphics[angle=-90,width=0.55\textwidth]{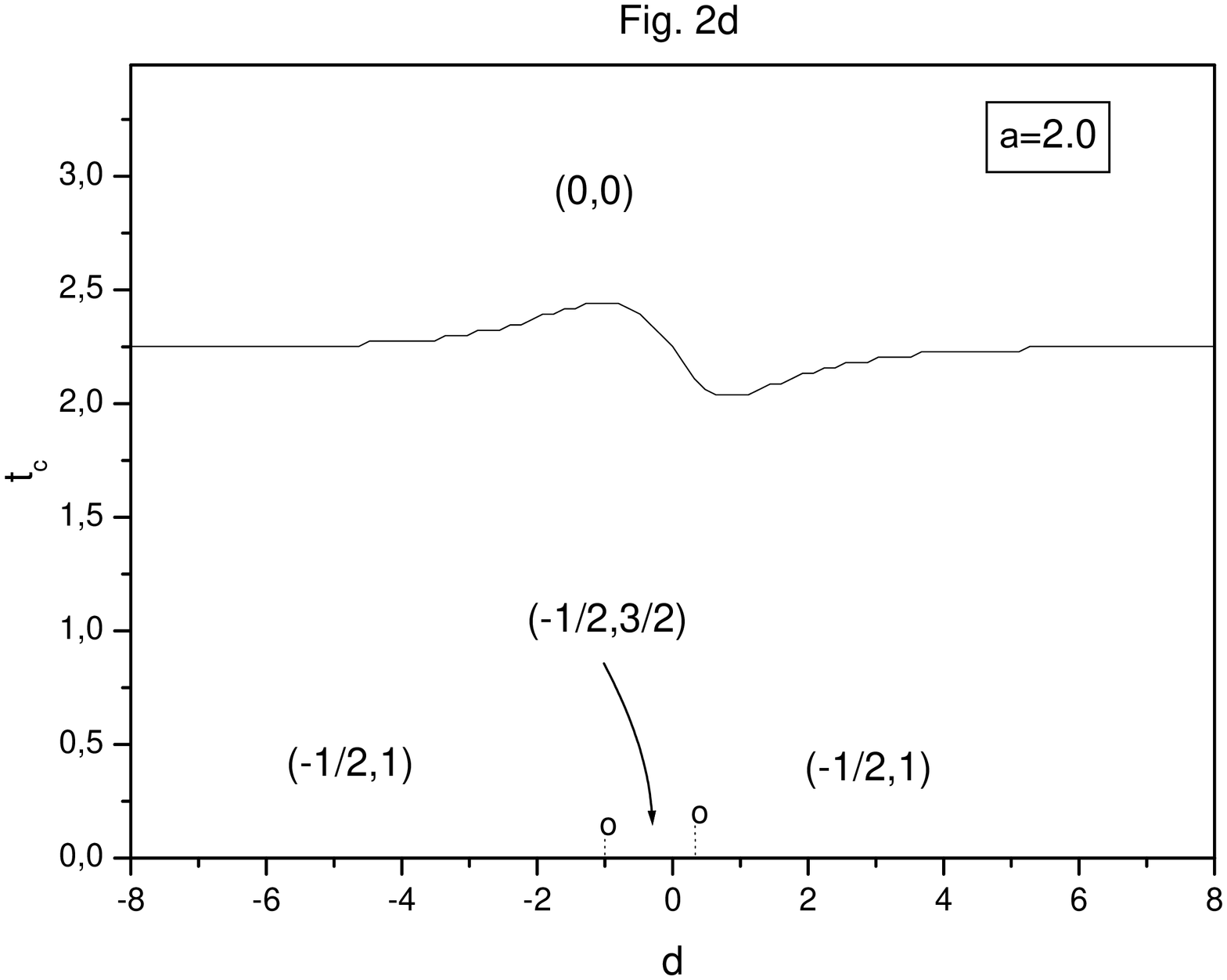}
\end{center}
\caption{The critical temperature as a function of $d$ plotted for: $\alpha =0$ (a), $\alpha =0.5$ (b), $\alpha =1$ (c) and $\alpha =2$ (d). The full lines correspond to the second-order transitions, whereas the dashed lines represent the first-order transitions. The tiny circles denote the isolated critical points.}
\end{figure}
In Fig. 2a ,$\alpha =0$, the paramagnetic $(0,0)$ and ferrimagnetic phases are separated by a second-order transition line (solid line). For very low temperatures, we found a first-order transition line (dashed line) separating the phases $(m_{\sigma}=\frac{-1}{2},m_S=\frac{3}{2})$ and $(m_{\sigma}=\frac{-1}{2},m_S=\frac{1}{2})$. This first-order line is terminated by an isolated critical point located at $(d_{tr}=1,t_{tr}=0.099)$. Whereas, for $\alpha =0.5$ see Fig. 2b, we found two first-order transition lines. In one hand, a first-order transition line separating the phases $(m_{\sigma}=\frac{-1}{2},m_S=\frac{3}{2})$ and $(m_{\sigma}=\frac{-1}{2},m_S=1)$, and terminated by an isolated critical point located at $(d_{tr}=2/3,t_{tr}=0.15)$. On the other hand, a first-order transition line separating the phases $(m_{\sigma}=\frac{-1}{2},m_S=1)$ and $(m_{\sigma}=\frac{-1}{2},m_S=\frac{1}{2})$, terminated by an isolated critical point located at $(d_{tr}=2,t_{tr}=0.045)$. Above these points, a continuous passage to the paramagnetic phase occurs. In Fig. 2c, plotted for $\alpha=1$, the phase $(m_{\sigma}=\frac{-1}{2},m_S=\frac{1}{2})$ disappears at low temperature and the only first-order transition line, present in this region, separates the phases $(m_{\sigma}=\frac{-1}{2},m_S=\frac{3}{2})$ and $(m_{\sigma}=\frac{-1}{2},m_S=1)$, and terminated by an isolated critical point located at $(d_{tr}=1/2,t_{tr}=0.034)$. Therefore, the second order transition-line temperatures are greater than those found for $\alpha=0$ and $\alpha=0.5$. This phenomenon is still present for a higher value of $\alpha$ (see Fig. 2d for $\alpha=2$). Indeed, the ferrimagnetic phase $(m_{\sigma}=\frac{-1}{2},m_S=1)$ emerges for large and positive values of the crystal-field. This is due to a competition between positive and negative values of the crystal-field which favours the ferrimagnetic phase. Consequently, the system exhibits two first-order transition lines, for very low temperatures, terminated by two isolated critical points: $(d=-1,t=0.11)$ and $(d=1/3,t=0.19)$ separating the phases $(m_{\sigma}=\frac{-1}{2},m_S=1)$ and $(m_{\sigma}=\frac{-1}{2},m_S=\frac{3}{2})$, and the phases $(m_{\sigma}=\frac{-1}{2},m_S=\frac{3}{2})$ and $(m_{\sigma}=\frac{-1}{2},m_S=1)$ respectively. \\
To give more details concerning the first-order transition lines, we have developed, at low temperature, the free energy and entropy of the system. Indeed, the free energy and entropy are plotted as function of $d$ for $t=0.035$ and two $\alpha$ values: $0.25$ and $1$, in figures $3a$ and $3b$, respectively. In accordance with Figs. $2b$ and $2c$, we observe a discontinuous change of  the free energy slope at first-order transition temperatures. Consequently the entropy is discontinuous at these temperatures (see inset of Figs. $3a$ and $3b$). For the second-order transition lines, we have plotted the total free energy and entropy, versus temperature, for $\alpha=0.25$ and $d=-3$ (see Fig. $3c$). We can remark that the entropy slope varies discontinuously at a second-order transition temperature. As consequence, the specific heat will exhibit a discontinuity at this transition temperature.    
\begin{figure}[!ht]
\begin{center}
\includegraphics[angle=-90,width=0.75\textwidth]{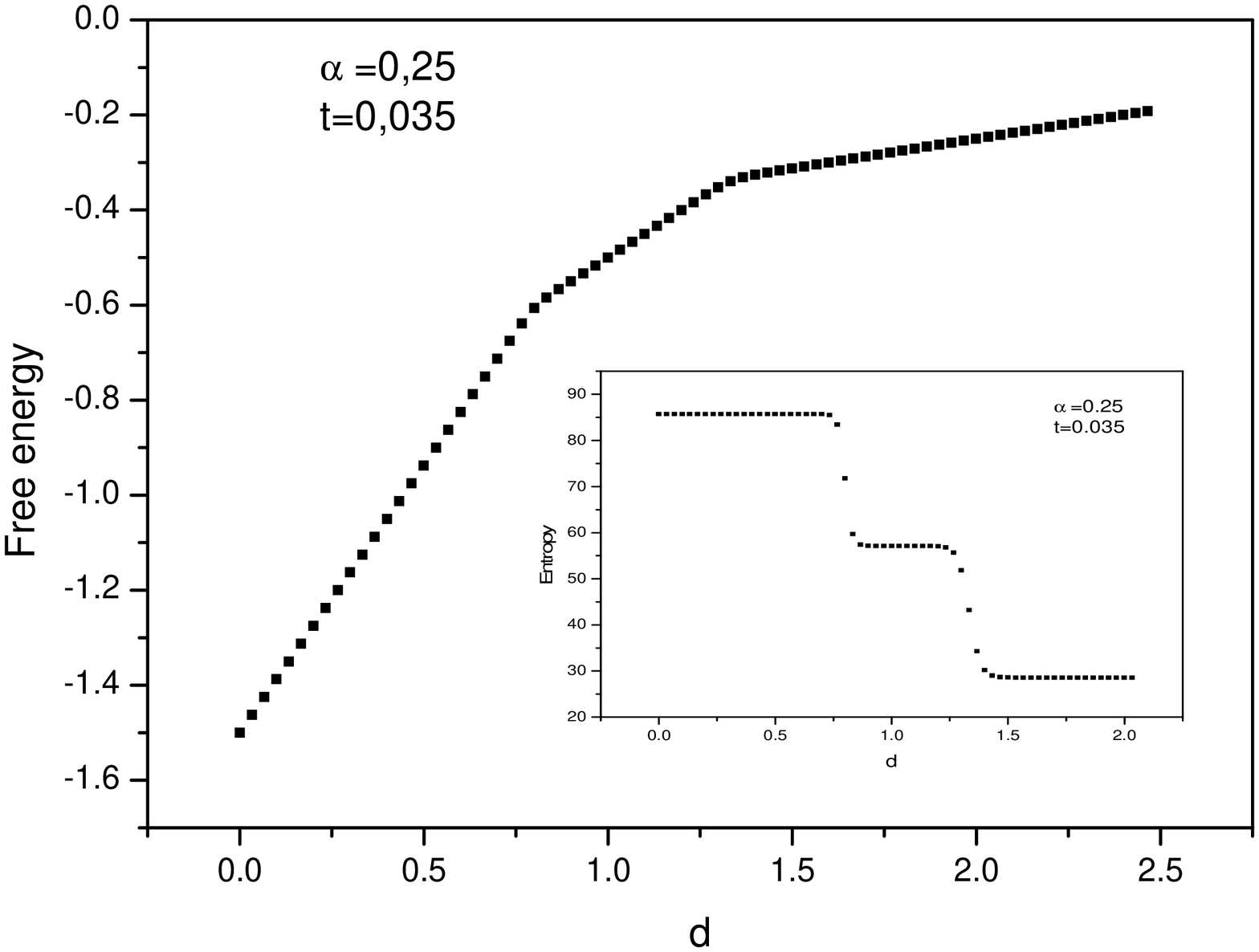}\qquad
\includegraphics[angle=-90,width=0.75\textwidth]{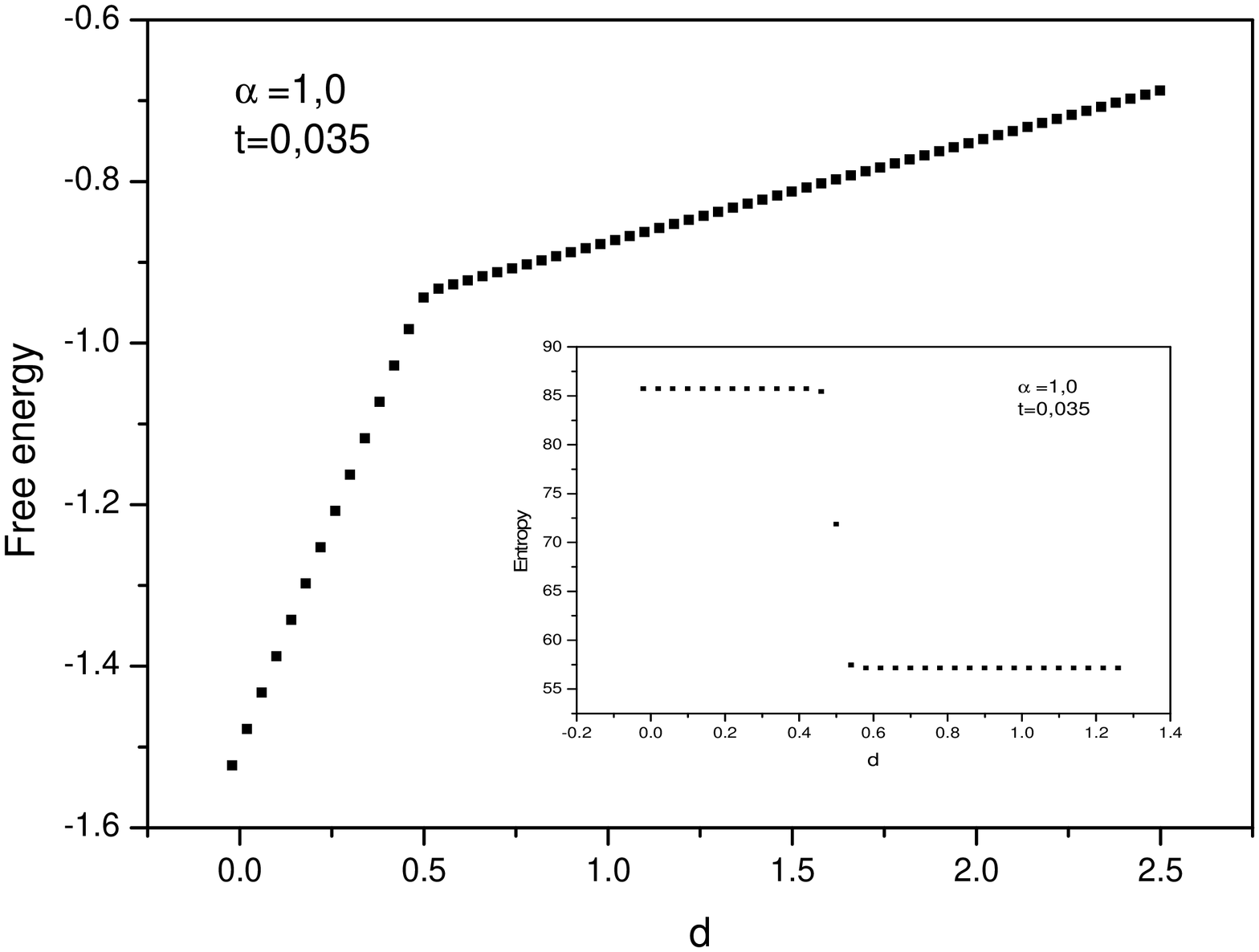}
\includegraphics[angle=-90,width=0.75\textwidth]{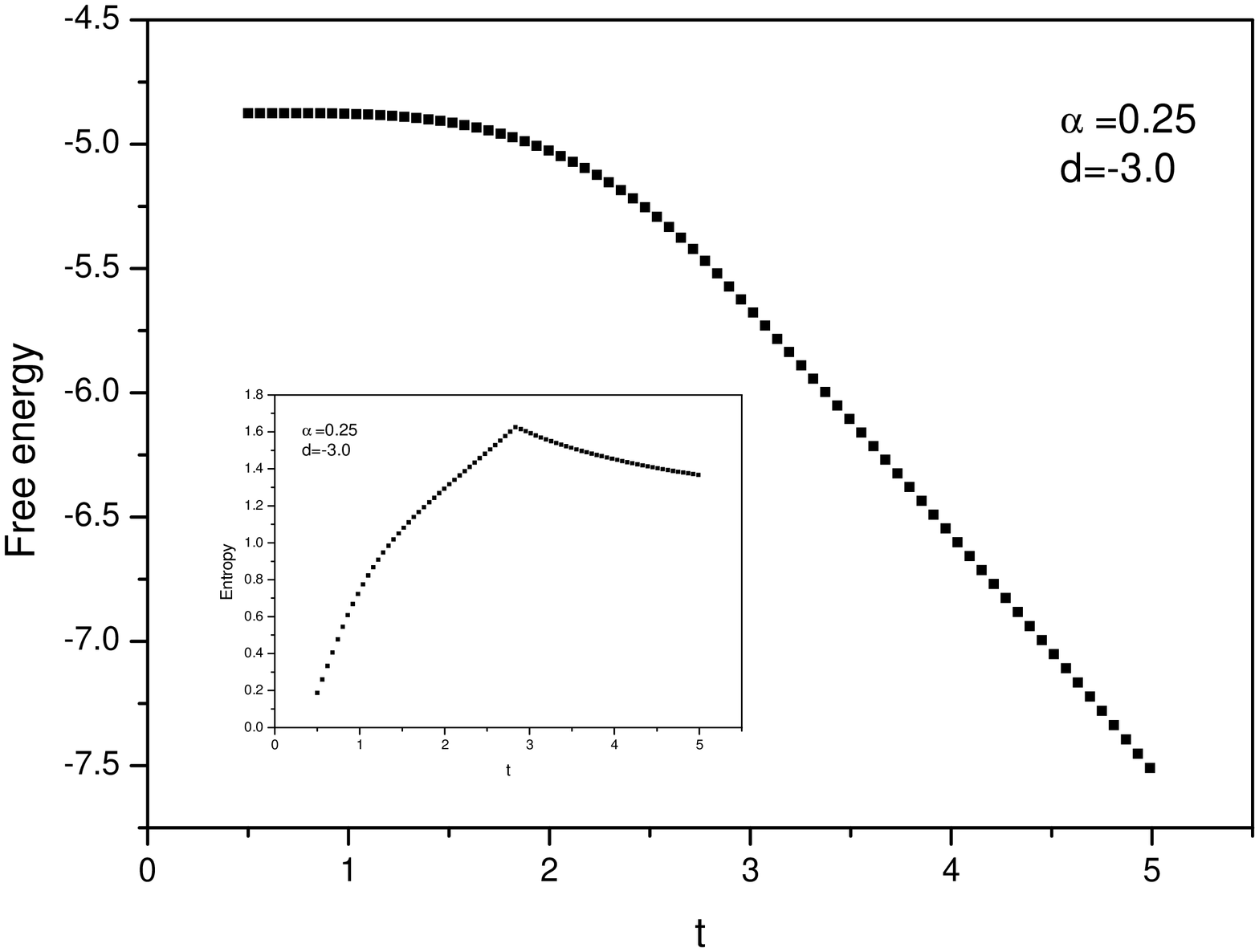}\qquad
\end{center}
\caption{In $a$ and $b$, the free energy and entropy (from expressions developed at low temperature) are plotted versus the reduced crystal-field $d$ for $t=0.035$ and two $\alpha$ values: $0.25$ and $1$, respectively. In $c$, the same physical quantities are plotted as a function of temperature for $\alpha = 0.25$ and $d=-3$.}
\end{figure}

In order to outline the effect of system behaviour as a function of the parameter $\alpha$ we plot, in Figs. 4, the transition temperatures $t_c$ for several values of the crystal field $d$. Indeed, for $d=2$, Fig. 4a shows the existence of a second-order transition line between the paramagnetic and the partly ferrimagnetic phases and a first-order transition between the phases $(m_{\sigma}=\frac{-1}{2},m_S=\frac{1}{2})$ and $(m_{\sigma}=\frac{-1}{2},m_S=1)$ terminated by an isolated critical point $(\alpha=3/2,t=0.055)$. Whereas for $d=0$ , see Fig. 4b, the second order transition line from the ferrimagnetic phase to the paramagnetic one is independent on $\alpha$ values because of the absence of the crystal field. In Fig. 4c, plotted for $d=-1$, we found a first order transition line separating the phases $(m_{\sigma}=\frac{-1}{2},m_S=\frac{3}{2})$ and $(m_{\sigma}=\frac{-1}{2},m_S=1)$, which terminates at an isolated critical point $(\alpha=2,t=0.09)$ between the ferrimagnetic and partly ferrimagnetic phases. At high temperatures, a second order transition line between these phases and the paramagnetic one is found.
\begin{figure}[!ht]
\begin{center}
\includegraphics[angle=-90,width=0.75\textwidth]{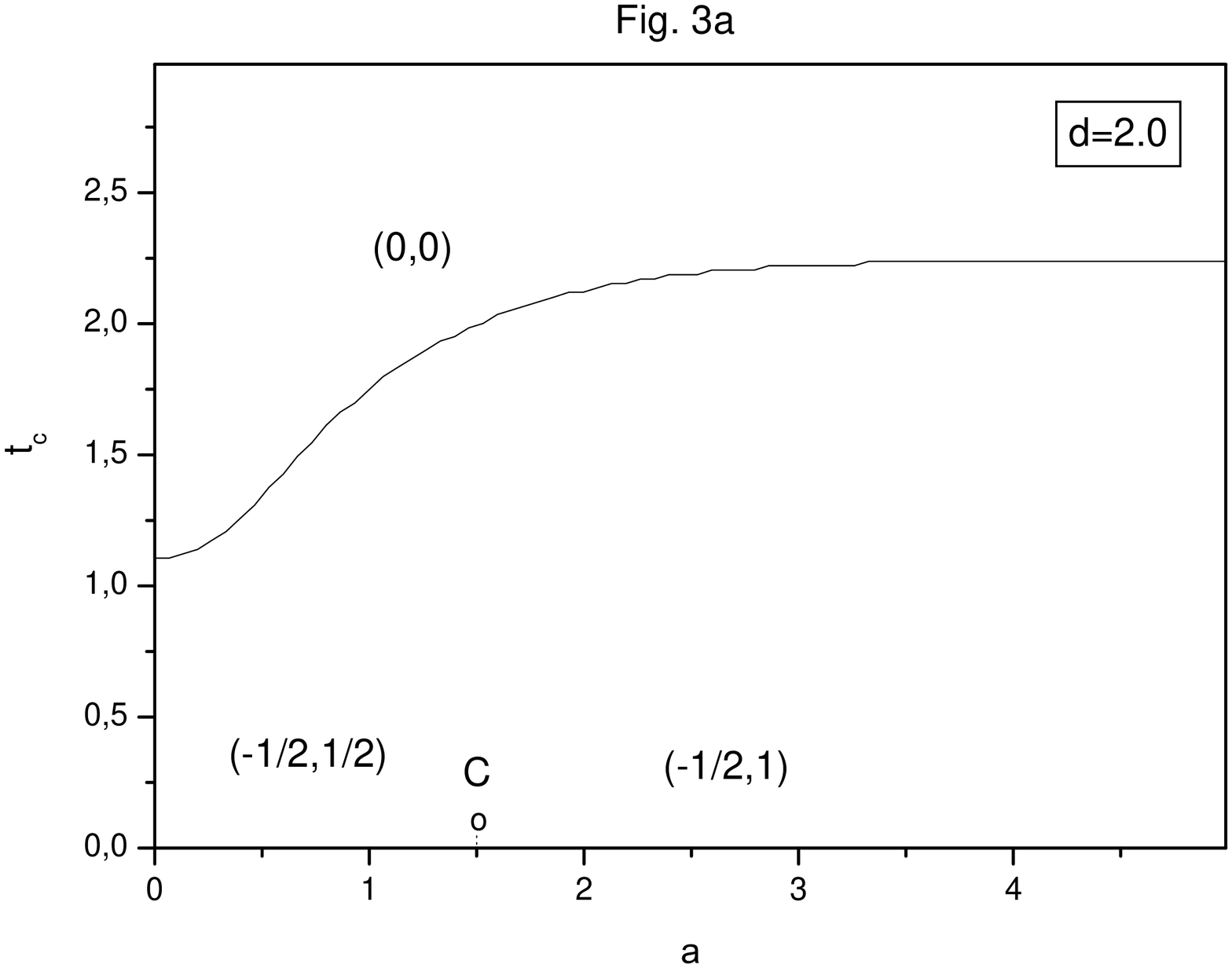}\qquad
\includegraphics[angle=-90,width=0.75\textwidth]{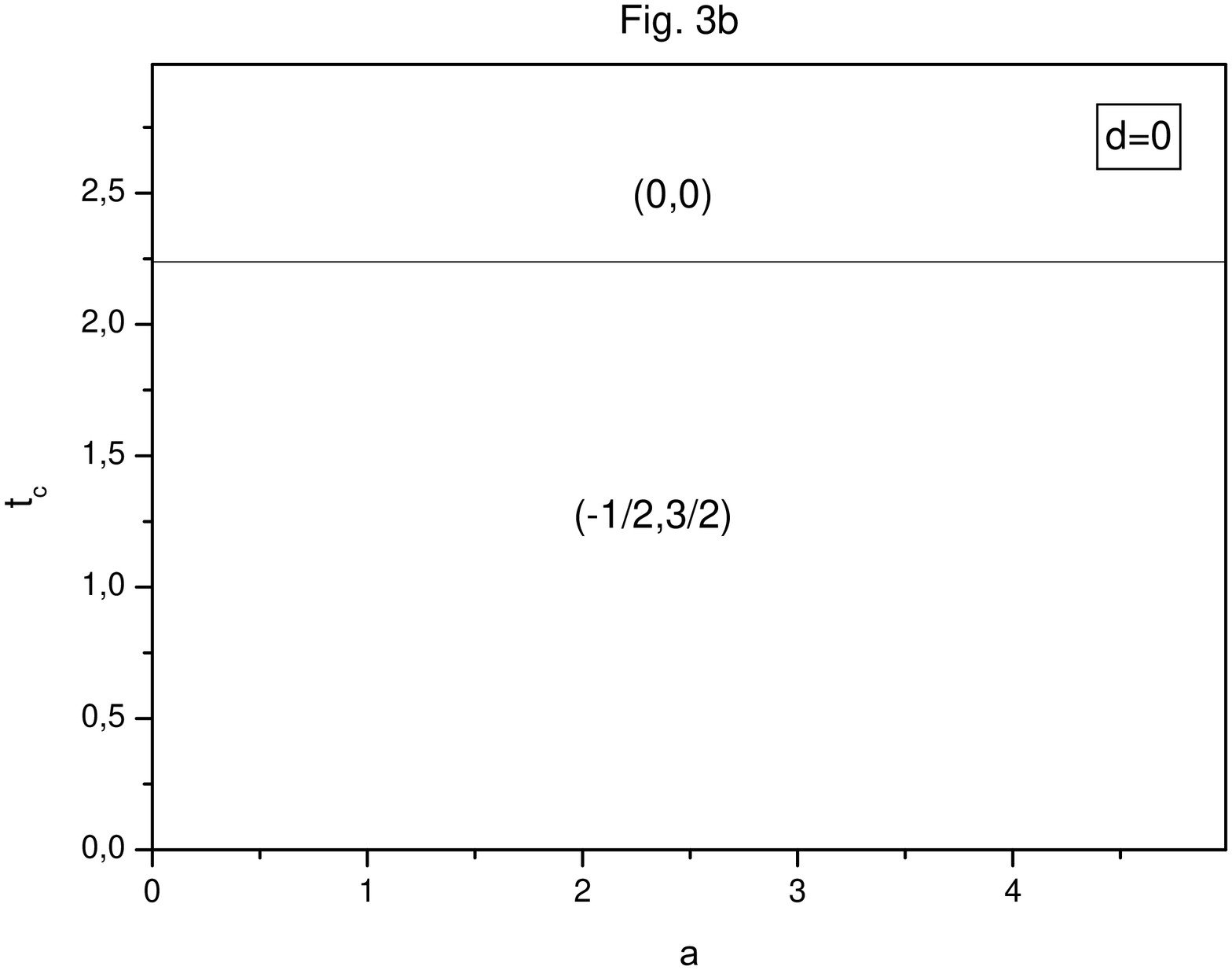}
\includegraphics[angle=-90,width=0.75\textwidth]{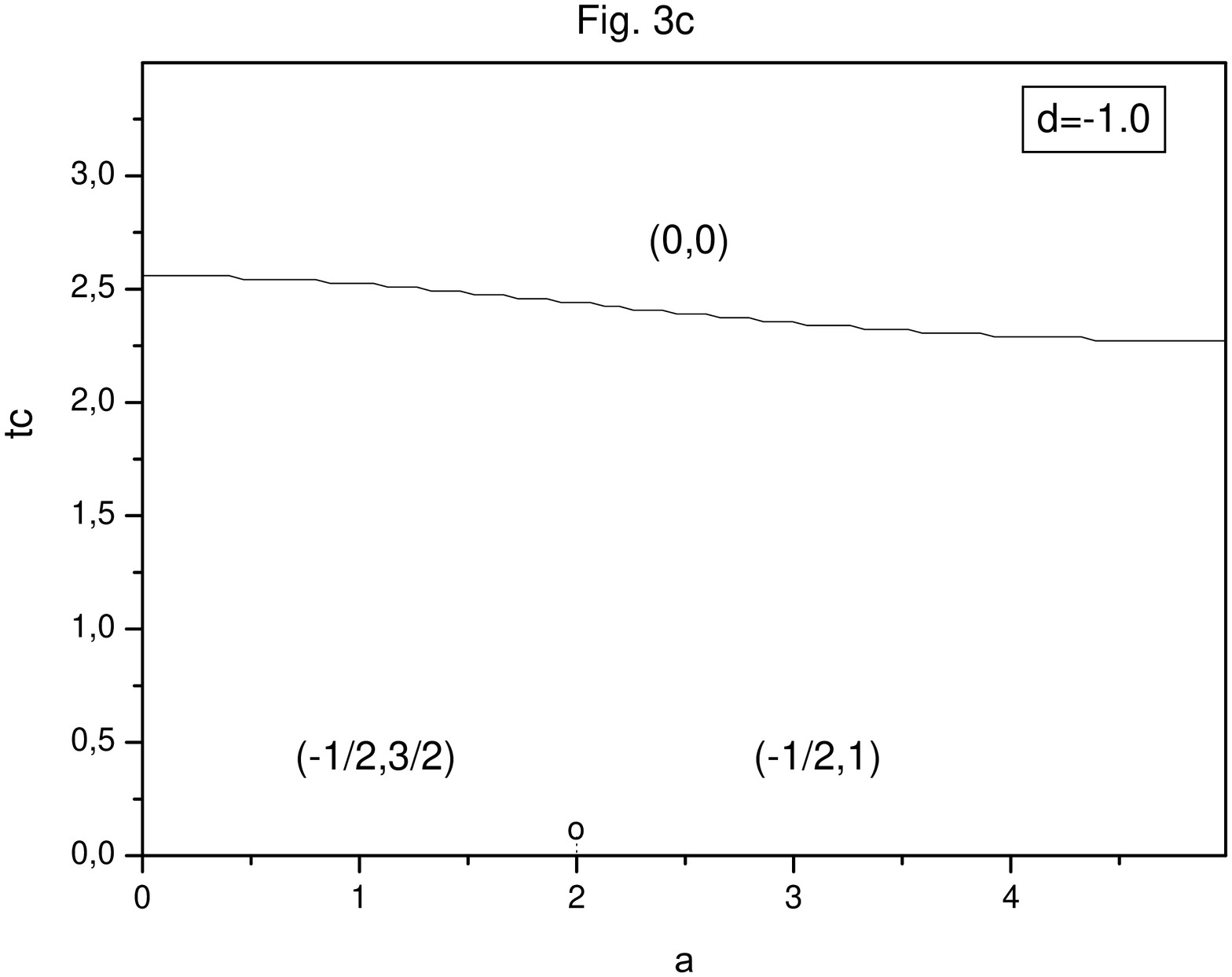}\qquad
\end{center}
\caption{The transition temperatures $t_c$ as a function of the parameter $\alpha$ for selected values of the crystal field: $d=2$ (a) , $d=0$  (b), and $d=-1$ (c).}
\end{figure}
\subsection{Magnetic properties}
In this part, we focus our interest on the magnetizations corresponding to the previous phase diagrams as a function of the crystal field $d$ and the parameter $\alpha$, for fixed temperature values. Indeed, Figs. 5a and 5b show the temperature dependence of magnetizations as a function of $d$ for $\alpha=0$ and $d=0.5$ respectively. The former figure presents a first order transition of the magnetization $m_S$, for a low temperature $t=0.05$: from $3/2$ to $1/2$ and a second order transition of the magnetization $m_S$, for a higher temperature $t=1.5$: from $3/2$ to the paramagnetic phase.
The second figure, Fig. 5b, shows a double first order transition of the magnetization $m_S$, for a very low temperature $t=0.04$: from $3/2$ to $1$ and from $1$ to $1/2$, and a second order transition for a higher temperature $t=1.1$: from $3/2$ to the paramagnetic phase for increasing values of the crystal field $d$.
In Fig. 5c we present the behaviour of the magnetization $m_S$ as a function of the crystal field for $\alpha=1$ and two temperature values: $t=0.03$ and $t=0.6$. In agreement with Fig. 2c, we found a discontinuity of this magnetization at a first order point between the phases $(m_{\sigma}=\frac{-1}{2},m_S=\frac{3}{2})$ and $(m_{\sigma}=\frac{-1}{2},m_S=1)$, for a low temperature ($t=0.03$). On the other hand, a continuous passage occurs from the same phases for a higher temperature ($t=0.6$).  \\
\begin{figure}[!ht]
\begin{center}
\includegraphics[angle=-90,width=0.75\textwidth]{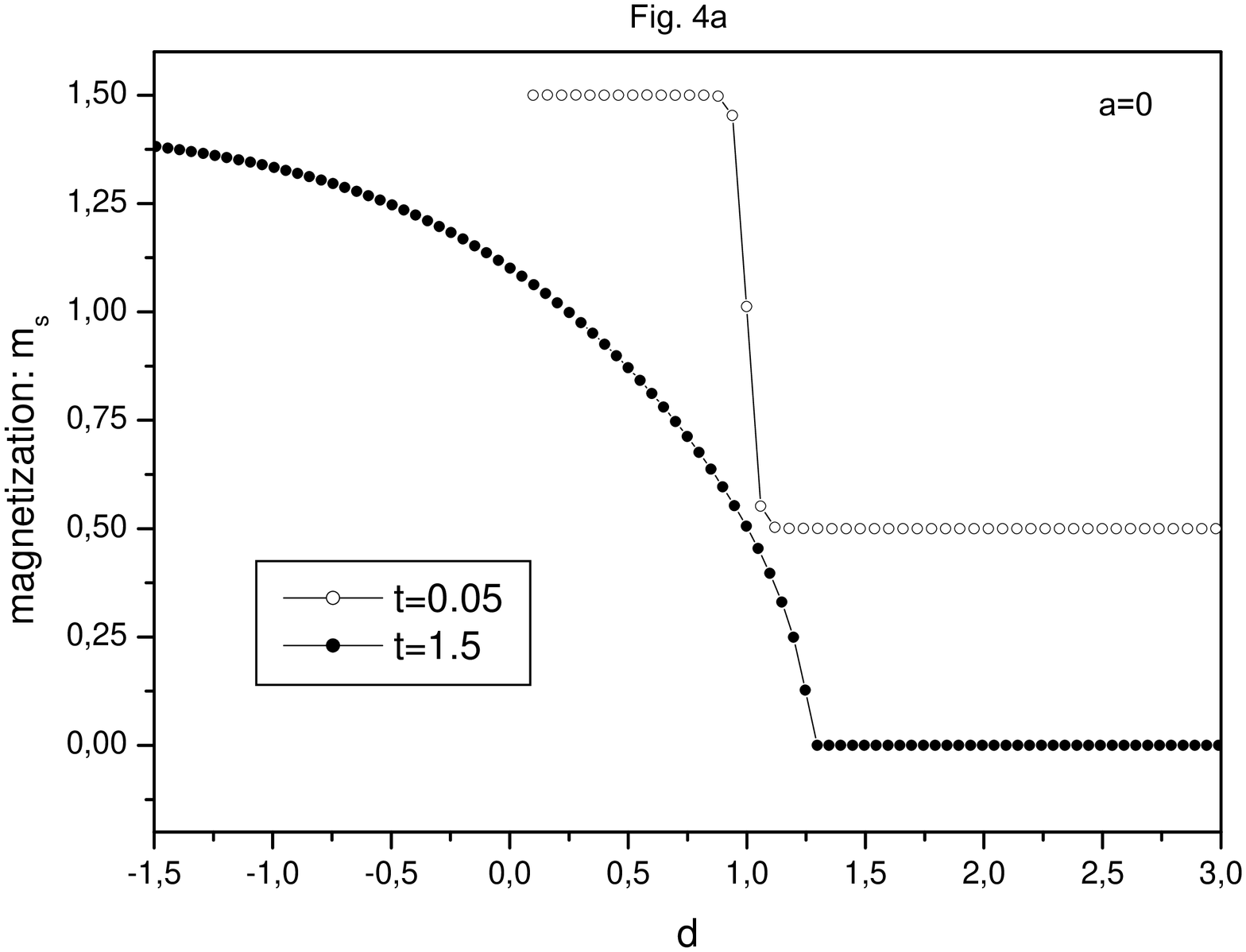}\qquad
\includegraphics[angle=-90,width=0.75\textwidth]{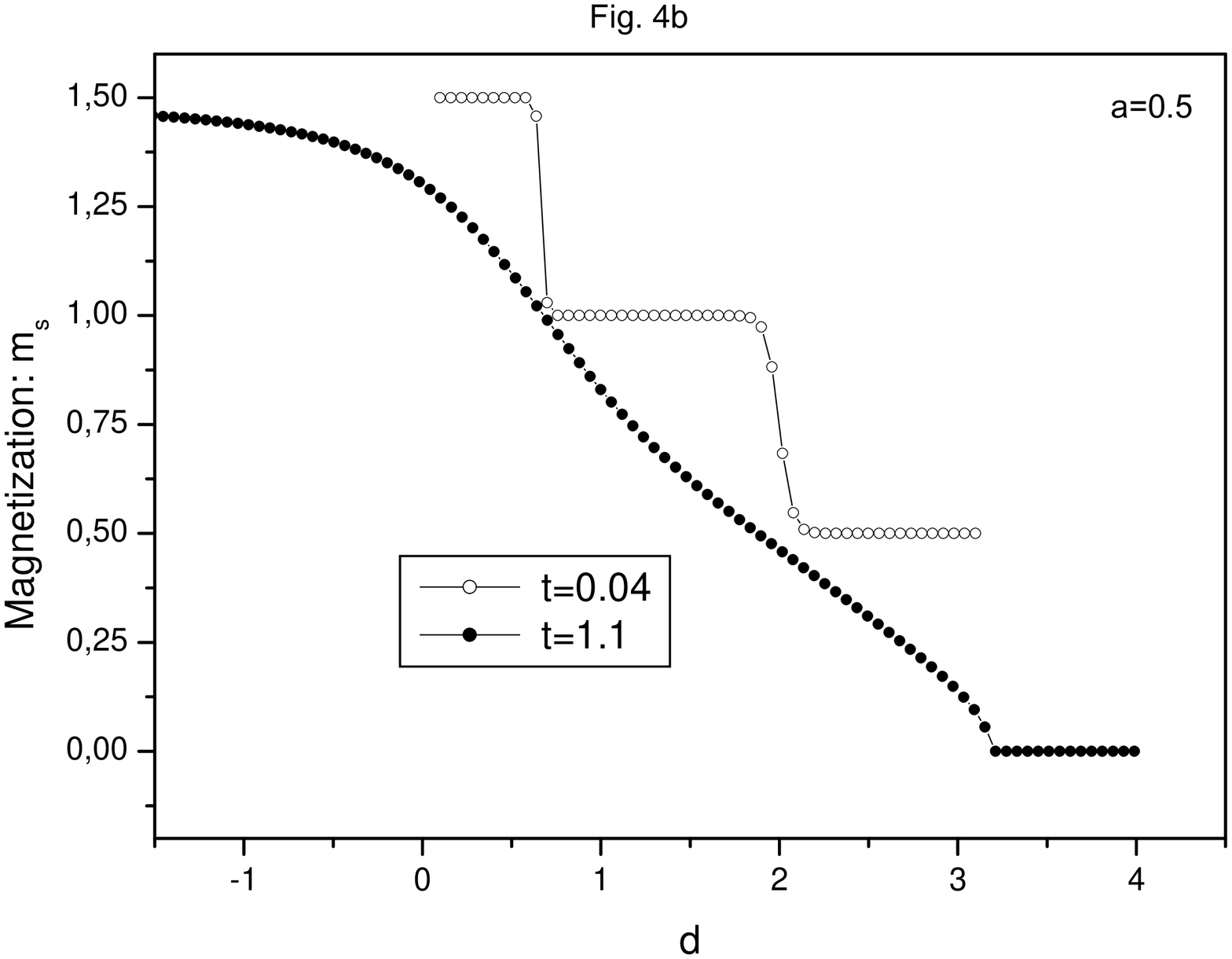}
\includegraphics[angle=-90,width=0.75\textwidth]{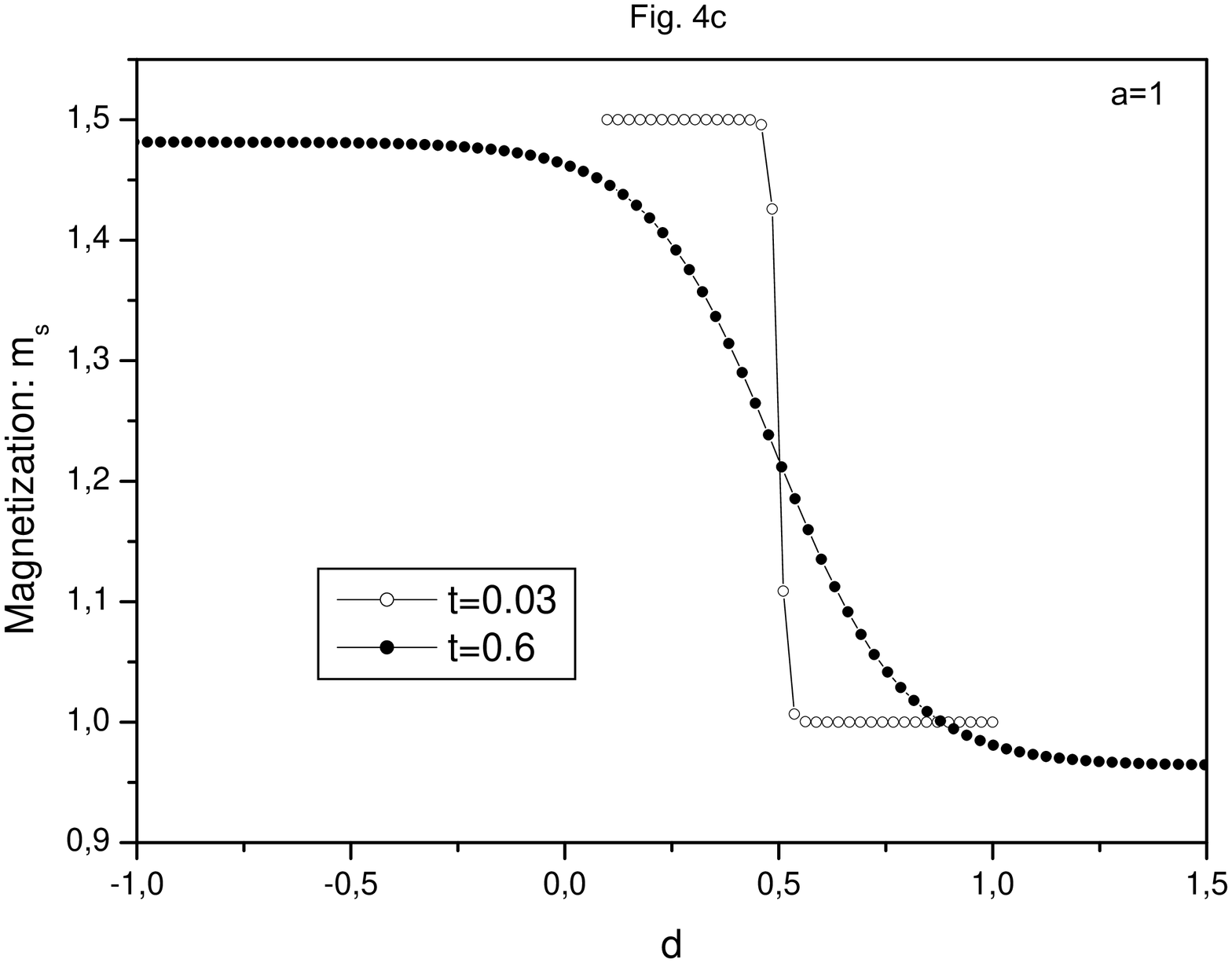}\qquad
\caption{The dependency of magnetization $m_S$ as a function of the reduced crystal-file $d$ for a fixed value of $\alpha$ and two reduced temperature values: $\alpha=0$ and ($t=0.05$ and $t=1.5$) (a), $\alpha=0.5$ and ($t=0.04$ and $t=1.1$) (b) and $\alpha=1.$ and ($t=0.03$ and $t=0.6$) (c), respectively..}
\end{center}
\end{figure}

The re-entrant behaviour found in Fig. 2d, is well illustrated in Figs. 6a and 6b plotted, for $\alpha=2$, and two temperatures $t=2.2$ and $t=2.3$, respectively. Indeed, Fig. 6a shows that for $t=2.2$, the magnetizations $m_{\sigma},m_S$ and consequently $M=(m_{\sigma}+m_S)/2$ drop to zero in the region $0 < d <3.5$. 
This is in good agreement with Fig. 2d. On the other hand, Fig. 6b shows that for a higher temperature ($t=2.3$) this phenomenon is inverted in the region $-3 < d <0$, so that the magnetizations $m_{\sigma},m_S$, and consequently $M=(m_{\sigma}+m_S)/2$, are nulls out of this region. This is once again in agreement with Fig. 2d. \\
\begin{figure}[!ht]
\begin{center}
\includegraphics[angle=-90,width=0.75\textwidth]{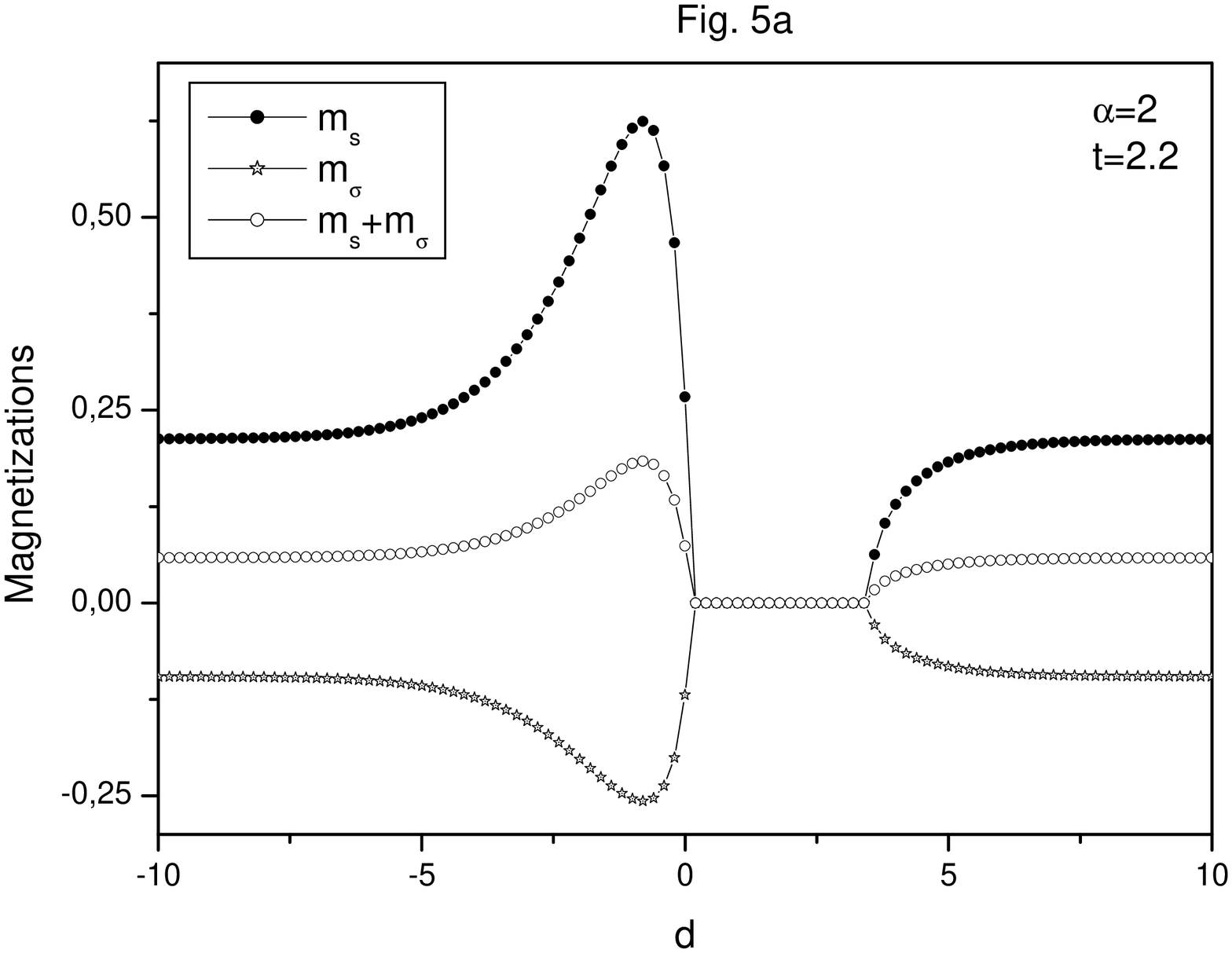}\qquad
\includegraphics[angle=-90,width=0.75\textwidth]{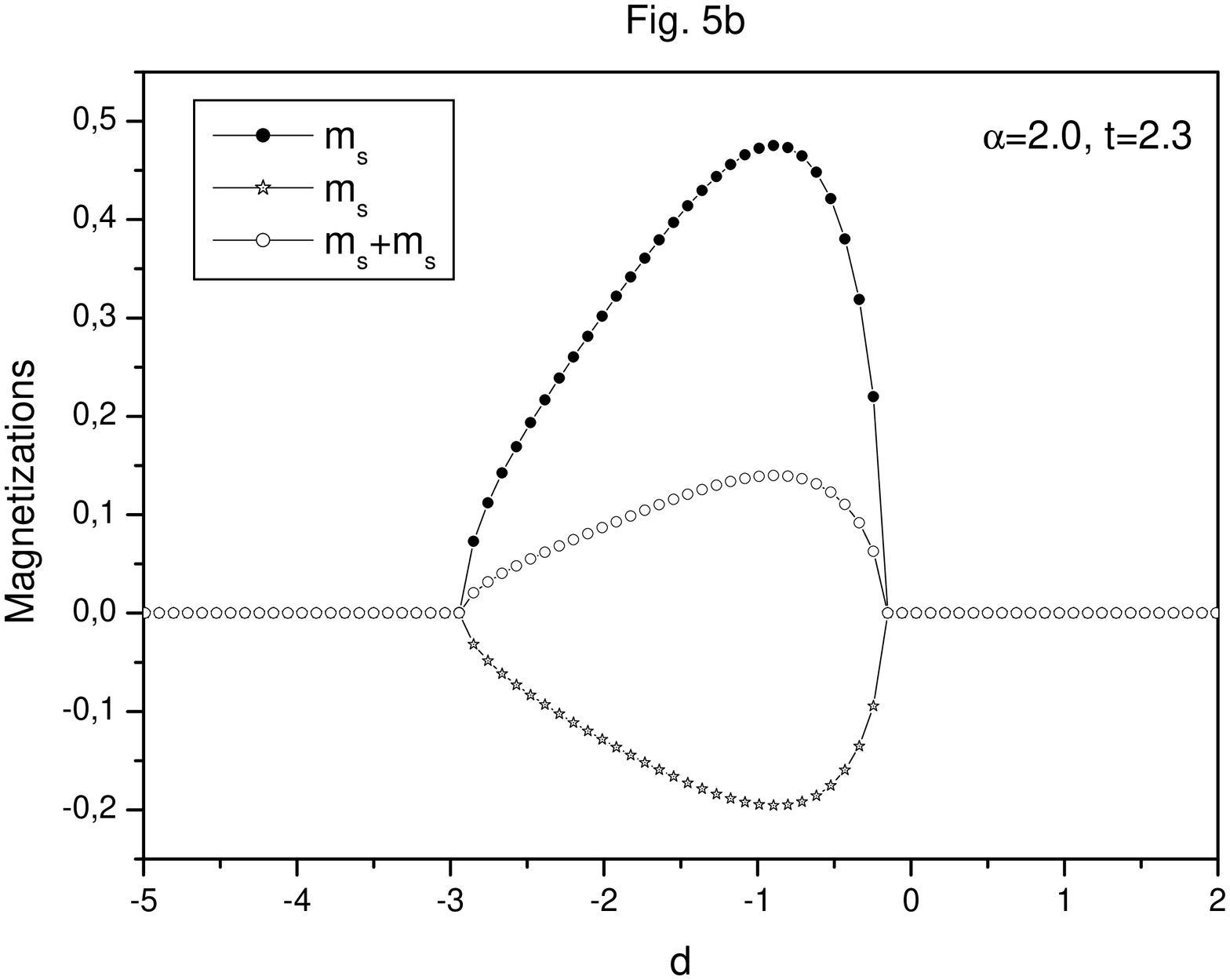}
\caption{The dependency of magnetizations $m_S$, $m_{\sigma}$ and $M=(m_{\sigma}+m_S)/2$ as a function of the reduced crystal-file $d$ for $\alpha=2$, and two temperature values: $t=2.2$ (a) and $t=2.3$ (b), respectively.}
\end{center}
\end{figure}
Let us now show the behaviour of the magnetizations $m_{\sigma},m_S$, and $M=(m_{\sigma}+m_S)/2$ as a function of the parameter $\alpha$ for fixed values of crystal field and temperature. For this purpose we plot in Figs. 7a and 7b these magnetizations as a function of $\alpha$ for $(d=2,t=2)$ and $(d=-1,t=2.5)$ respectively. The first figure shows that for positive values of the crystal field, the increasing $\alpha$ values effect is to force the ordered phase, in a region of temperatures lower than $t_c$. This is the case in Fig. 7a, for $t=2$. For a negative value of the crystal field, this phenomenon is inverted as it is shown in Fig. 7b, plotted for $d=-1$. One can note that the results found in Figs. 7a and 7b, are in good agreement with those illustrated in Figs. 4a and 4c, respectively. On the other hand, it is obvious that the magnetizations are independent of the parameter $\alpha$ in absence of the crystal field, as it is well illustrated in Fig. 4b. \\
\begin{figure}[!ht]
\begin{center}
\includegraphics[angle=-90,width=0.75\textwidth]{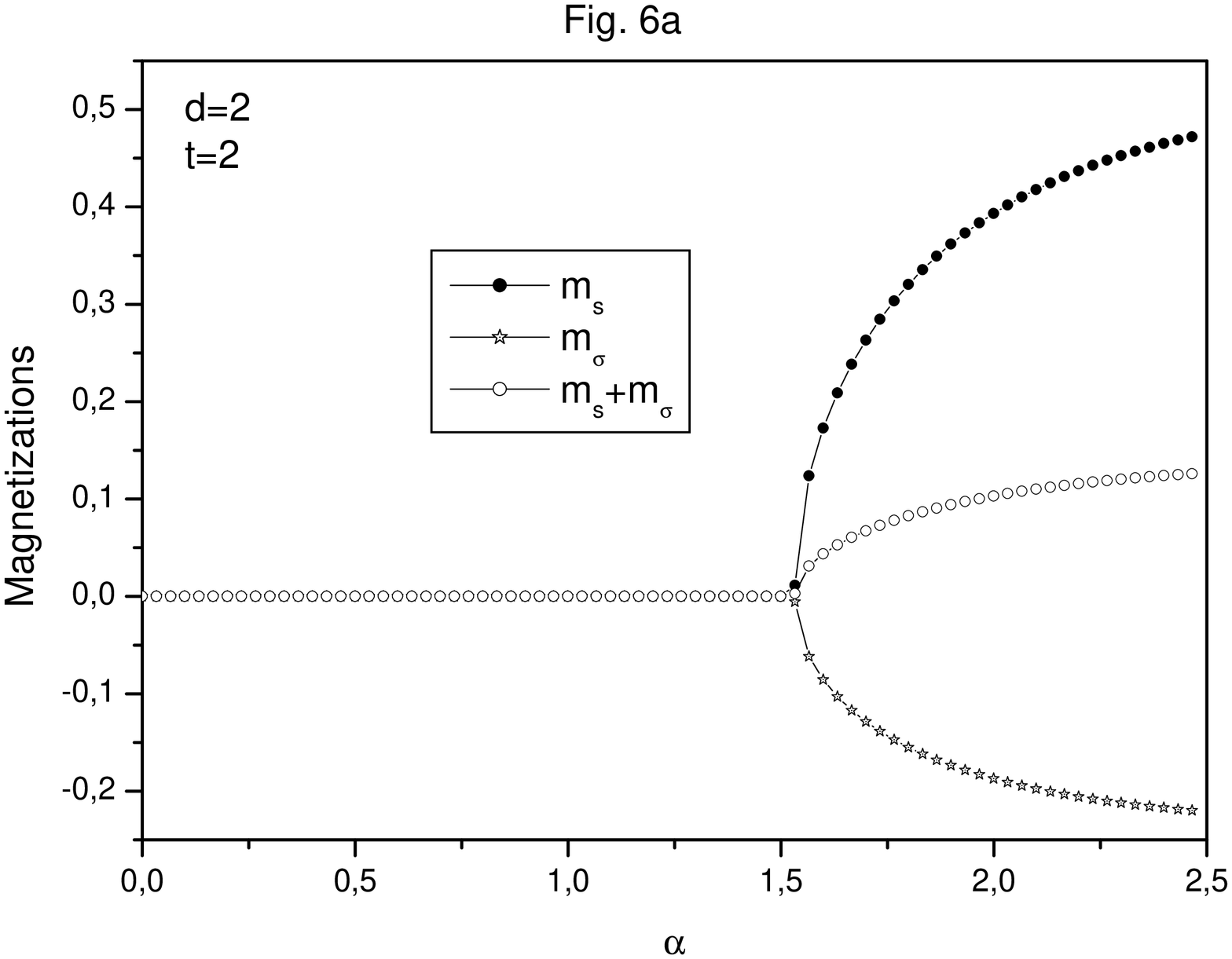}\qquad
\includegraphics[angle=-90,width=0.75\textwidth]{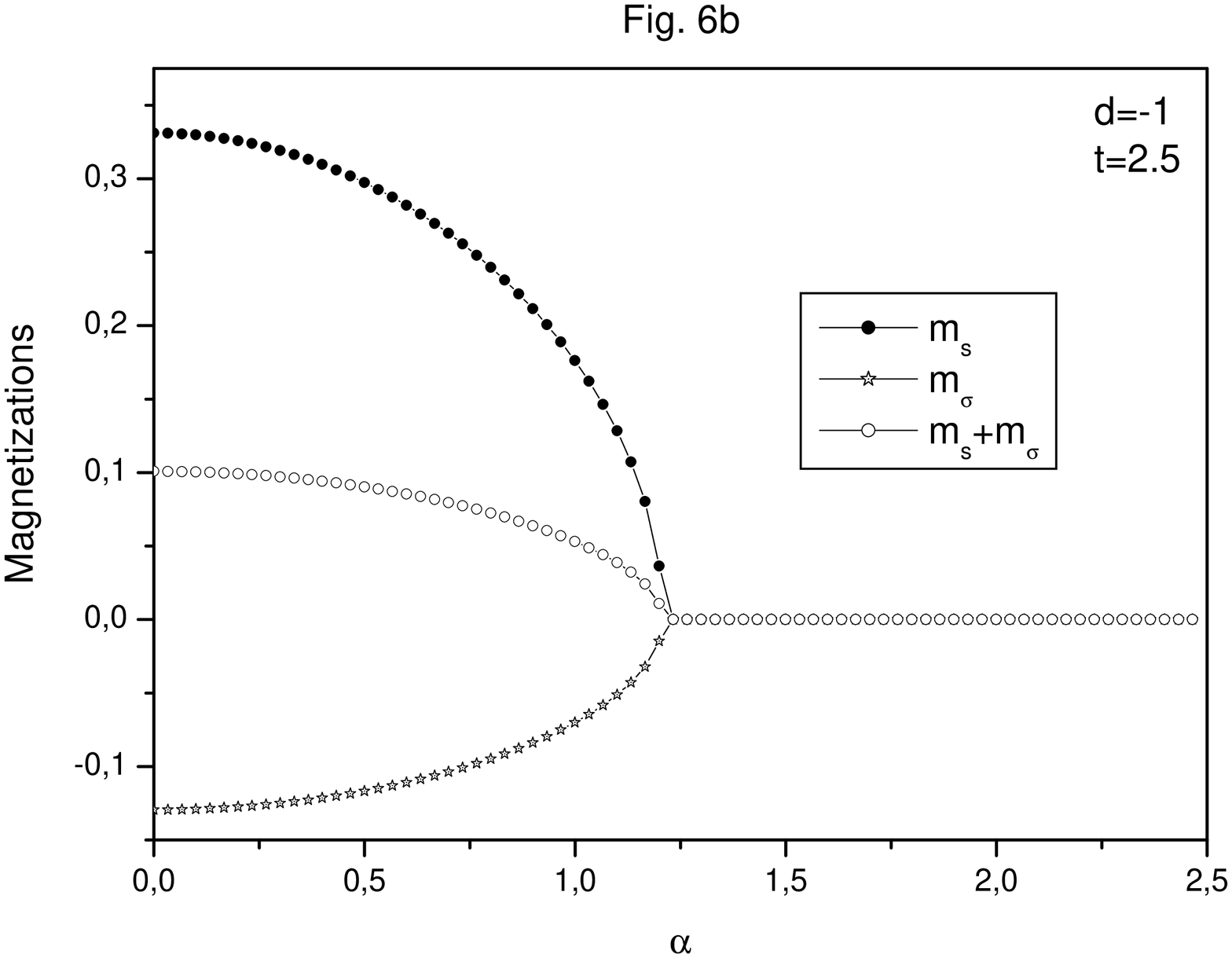}
\end{center}
\caption{The behaviour of the magnetizations $m_{\sigma},m_S$, and $M=(m_{\sigma}+m_S)/2$ as a function of the parameter $\alpha$ for fixed values crystal field and temperature for $(d=2,t=2)$ in (a) and $(d=-1,t=2.5)$ in (b). }
\end{figure}
To complete this study, we have investigated the effect of increasing the exchange interaction parameter $J$, at very low temperature when keeping $\alpha$ constant. Indeed, the results we found for a low temperature $t=0.025$ and selected values of $\alpha$, showed that the transitions obtained in the ground state phase diagram (Fig. 1) are still present. For $\alpha=0.5$, Fig. 8a showed that the system can exhibit the phases $(m_{\sigma}=\frac{-1}{2},m_S=\frac{1}{2})$, $(m_{\sigma}=\frac{-1}{2},m_S=1)$ and $(m_{\sigma}=\frac{-1}{2},m_S=\frac{3}{2})$ when increasing the parameter $J$ at a positive and constant crystal field ($d > 0$). For $\alpha=1.0$, the system undergoes a transition from the phase $(m_{\sigma}=\frac{-1}{2},m_S=1)$ to the phase $(m_{\sigma}=\frac{-1}{2},m_S=\frac{3}{2})$
for positive and constant crystal field, when increasing the exchange interaction $J$, see Fig. 8b. For a large value of $\alpha$ ($\alpha=2$), Fig. 8c shows that the effect of increasing the parameter $J$ on the phase transitions $(m_{\sigma}=\frac{-1}{2},m_S=\frac{1}{2})$, $(m_{\sigma}=\frac{-1}{2},m_S=1)$ and $(m_{\sigma}=\frac{-1}{2},m_S=\frac{3}{2})$ is to displace these transitions towards large and positive values of the crystal field.
\newpage
\begin{figure}[!ht]
\begin{center}
\includegraphics[angle=-90,width=0.75\textwidth]{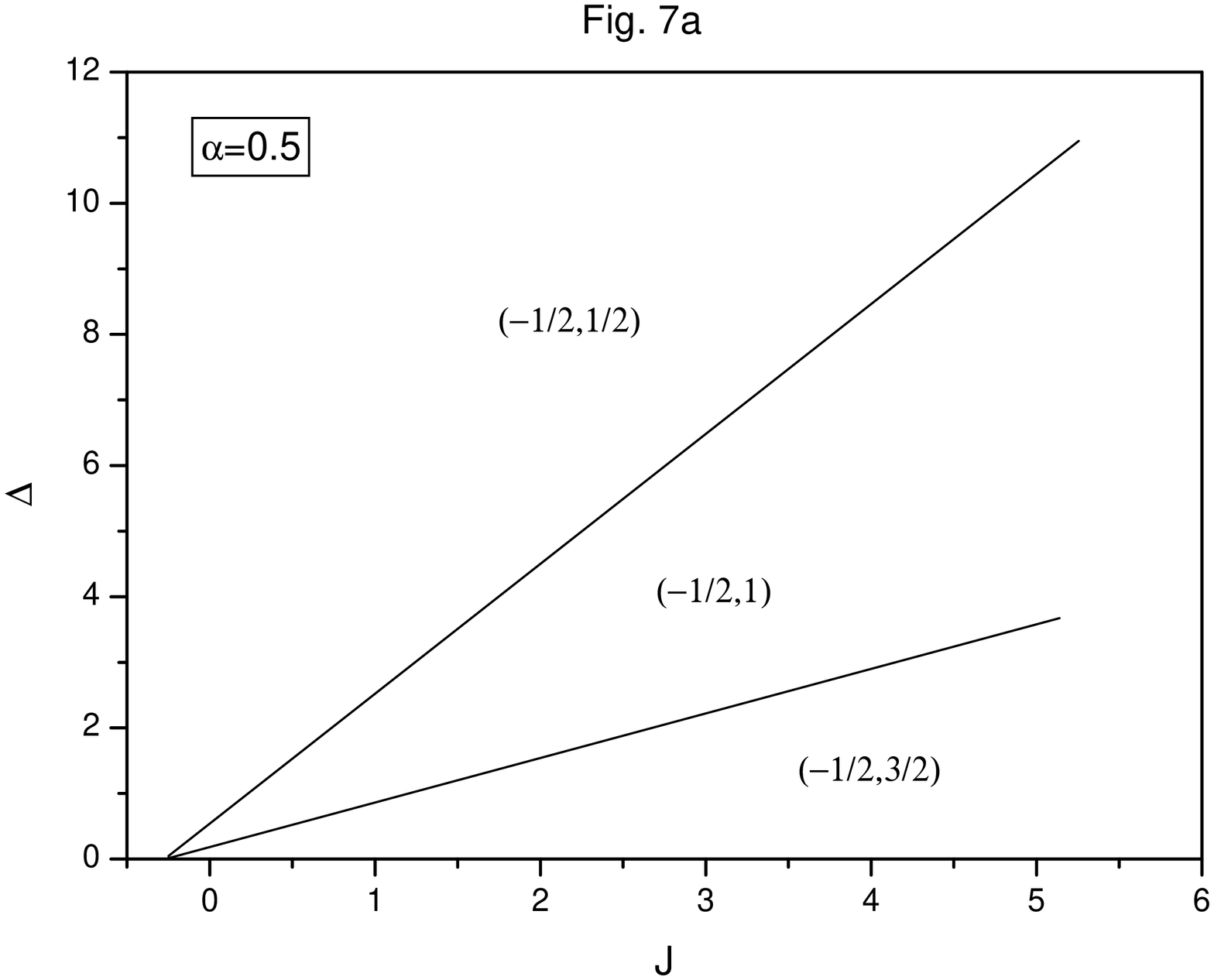}\qquad
\includegraphics[angle=-90,width=0.75\textwidth]{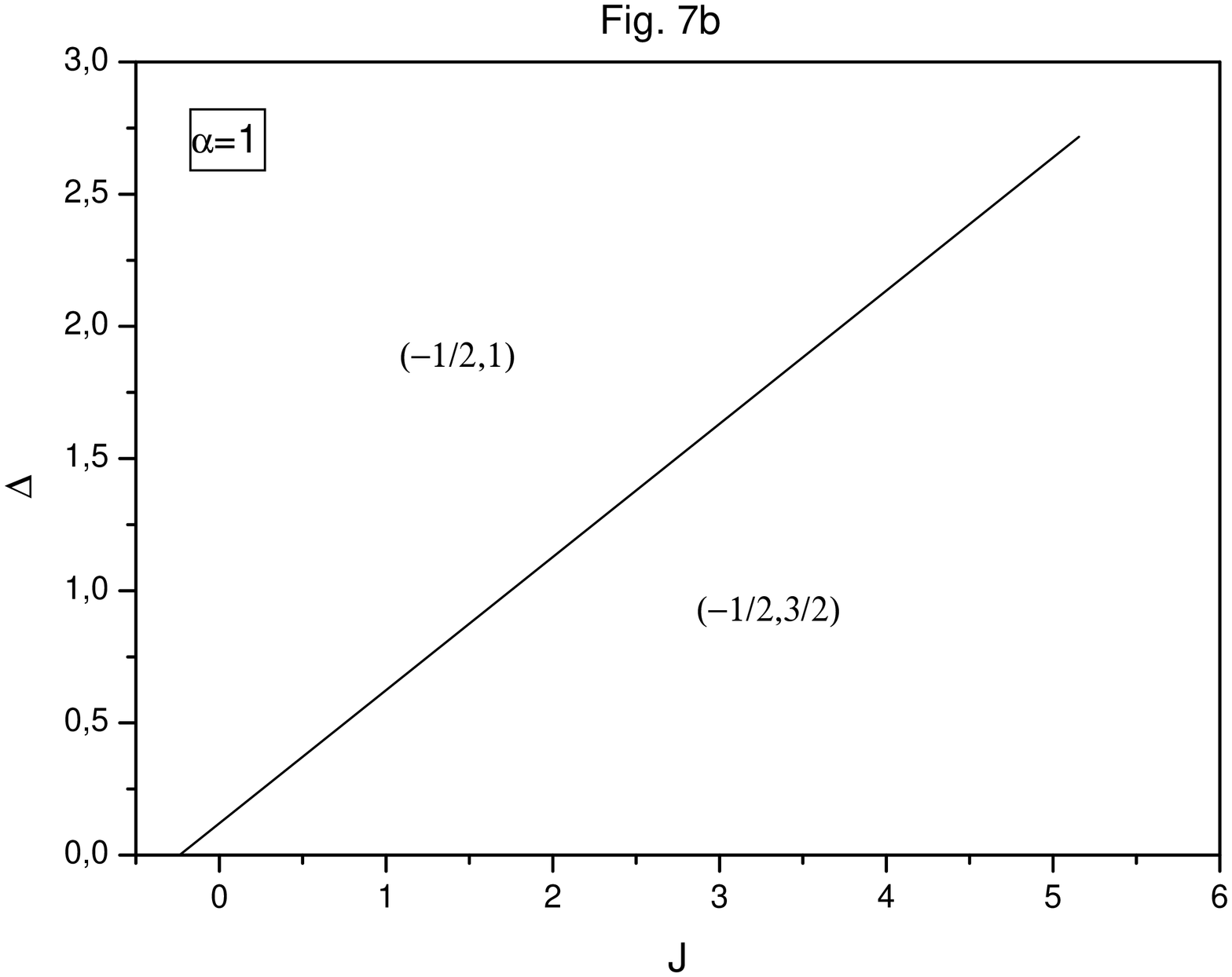}
\includegraphics[angle=-90,width=0.75\textwidth]{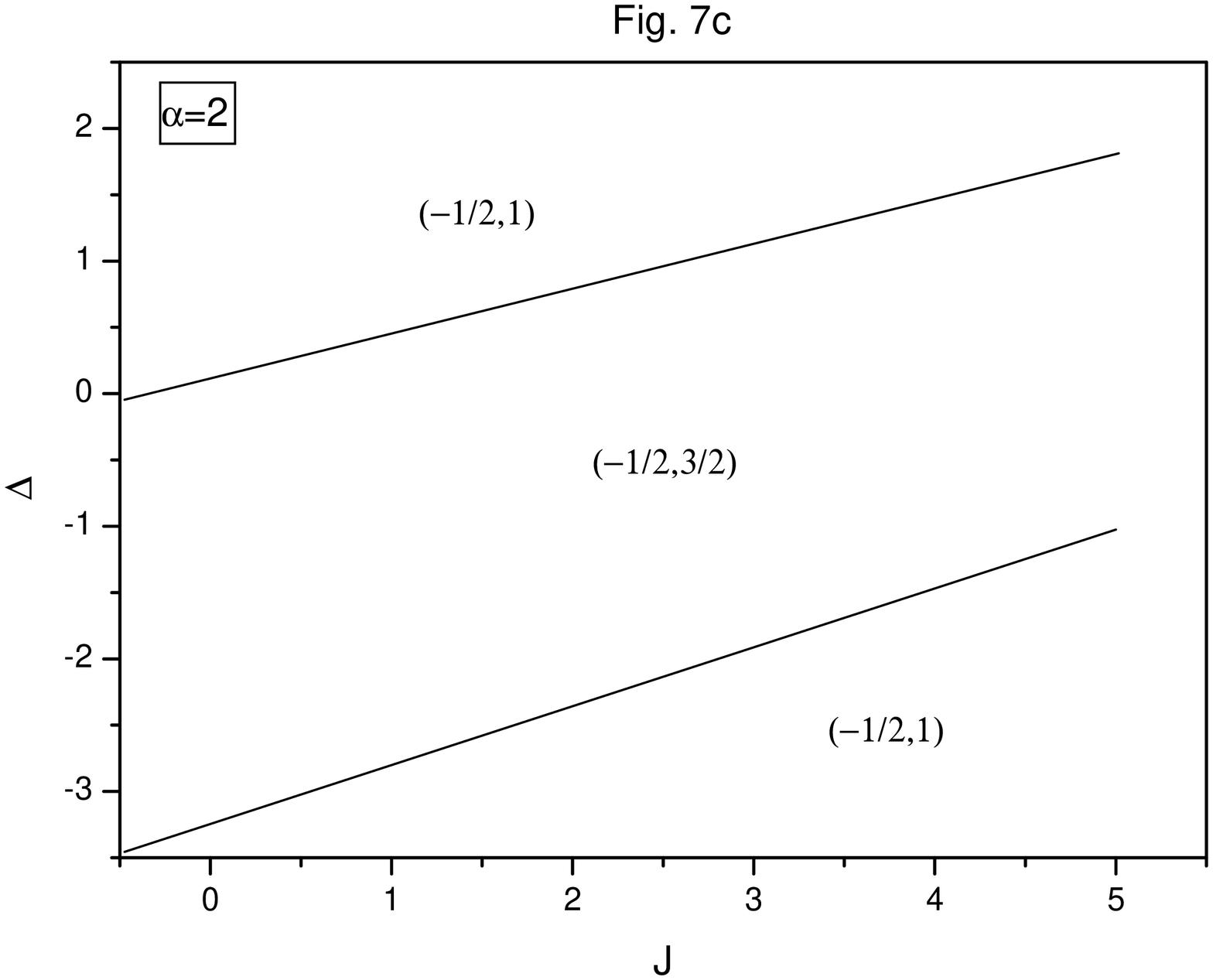}\qquad
\end{center}
\caption{Phase diagrams in the plane $(\Delta,J)$, at very low temperature $t=0.025$ for $\alpha=0.5$ (a), $\alpha=1$ (b) and $\alpha=2$ (c). In (a)  the system exhibits the phases $(m_{\sigma}=\frac{-1}{2},m_S=\frac{1}{2})$, $(m_{\sigma}=\frac{-1}{2},m_S=1)$ and $(m_{\sigma}=\frac{-1}{2},m_S=\frac{3}{2})$, in (b) the system undergoes a transition from the phase $(m_{\sigma}=\frac{-1}{2},m_S=1)$ to the phase $(m_{\sigma}=\frac{-1}{2},m_S=\frac{3}{2})$, while (c) shows the phases $(m_{\sigma}=\frac{-1}{2},m_S=\frac{1}{2})$, $(m_{\sigma}=\frac{-1}{2},m_S=1)$ and $(m_{\sigma}=\frac{-1}{2},m_S=\frac{3}{2})$ when increasing the exchange interaction parameter $J$ for fixed values of the crystal field $\Delta$.}
\end{figure}

\section{Conclusions}
We have investigated a mixed spin $\sigma=1/2$ and spin $S=3/2$ Ising model on a square lattice, using the mean field approximation $(MFA)$. The effect of a random crystal field on the magnetic properties of the system is investigated. Indeed, our results revealed many interesting phenomena, namely, several topologically different types of phase diagrams. Furthermore, these phase diagrams present rich varieties of phase transitions with first and second order phase transition lines. These lines are linked by tricritical points and terminated at isolated critical points. Finally, the effect of increasing the exchange interaction parameter $J$, at very low temperature when keeping $\alpha$ constant, is investigated. The results we found for a very low temperature and sleeted values of $\alpha$, showed that the transitions obtained in the ground state phase diagram are still present, but displaced towards large and positive values of the crystal field. \\


\end{document}